\documentstyle[12pt]{article}

\newcommand{\be}{\begin{equation}}
\newcommand{\ee}{\end{equation}}
\newcommand{\bea}{\begin{eqnarray}}
\newcommand{\eea}{\end{eqnarray}}
\newcommand{\nn}{\nonumber \\}
\newcommand{\p}[1]{(\ref{#1})}
\newcommand{\ba}{\begin{array}}
\newcommand{\ea}{\end{array}}
\newcommand{\vs}[1]{\vspace{#1 mm}}

\renewcommand{\a}{\alpha}
\renewcommand{\b}{\beta}
\renewcommand{\c}{\gamma}
\renewcommand{\d}{\delta}
\newcommand{\e}{\epsilon}

\newcommand{\tOmega}{{\widetilde\Omega}}
\newcommand{\tpsi}{{\tilde\psi}}
\newcommand{\tsigma}{{\tilde\sigma}}
\newcommand{\ts}{{\tilde\sigma}}
\newcommand{\tw}{{\tilde w}}
\newcommand{\te}{{\tilde e}}
\newcommand{\bQ}{{\widetilde Q}}

\def\tL{{\widetilde L}}
\def\tB{{\widetilde B}}
\def\tW{{\widetilde W}}

\def\bbox{{\,\lower0.9pt\vbox{\hrule \hbox{\vrule height 0.2 cm
\hskip 0.2 cm \vrule height 0.2 cm}\hrule}\,}}
\newcommand{\dsl}{\pa \kern-0.5em /}

\newcommand{\pa}{\partial}

\font\mybb=msbm10 at 12pt
\def\bb#1{\hbox{\mybb#1}}

\def\bR {\bb{R}}
\def\bE {\bb{E}}

\def\a{\alpha}\def\b{\beta}\def\g{\gamma}
\def\d{\delta}

\def\cosec{\,{\rm cosec}\,}
\def\ie{{\it i.e.,}\ }
\def\eg{{\it e.g.,}\ }

\begin{document}

\topmargin 0pt
\oddsidemargin 5mm

\renewcommand{\thefootnote}{\fnsymbol{footnote}}
\begin{titlepage}

\setcounter{page}{0}

\rightline{\small hep-th/9809065 \hfill UCSB NSF-ITP-98-075}
\vskip -.75em\rightline{\small \hfill McGill/98-12}
\vskip -.75em\rightline{\small \hfill QMW-PH-98-27}

\vs{15}
\begin{center}
{\Large SUPERSYMMETRY OF ROTATING BRANES}
\vs{10}

{\large
J.P. Gauntlett$^1$, R.C. Myers$^{2,}$\footnote{Permanent address: Department
 of Physics, McGill University, Montr\'eal, PQ, H3A 2T8, Canada}
 and  P.K. Townsend$^{2,}$\footnote{Permanent address: 
DAMTP, University of Cambridge, Cambridge CB3 9EW, U.K.} } \\
\vs{5}
${}^1${\em Department of Physics\\
       Queen Mary and Westfield College\\
       University of London\\
       Mile End Road\\
       London E1 4NS, UK}\\[2em]
${}^2${\em Institute for Theoretical Physics\\
University of California at Santa Barbara\\ 
CA 93106, USA.}
\end{center}
\vs{7}
\centerline{{\bf Abstract}}

We present a new 1/8 supersymmetric intersecting M-brane solution of D=11
supergravity with two independent rotation parameters. The metric has a
non-singular event horizon and the near-horizon geometry is 
$adS_3\times S^3\times S^3\times\bE^2$ (just
as in the non-rotating case). We also present a method of determining
the isometry supergroup of supergravity solutions from the Killing spinors and use it to
show that for the near horizon solution it is
$D(2|1,\alpha)\times D(2|1,\alpha)$ where
$\alpha$ is the ratio of the two 3-sphere radii.
We also consider various dimensional
reductions of our solution, and the corresponding effect of these
reductions on the Killing spinors and the isometry supergroups.

\end{titlepage}
\newpage
\renewcommand{\thefootnote}{\arabic{footnote}}
\setcounter{footnote}{0}

\section{Introduction}

Many of the $adS_{p+2}\times S^{D-p-2}$  Kaluza-Klein (KK) vacua of
$D$-dimensional supergravity theories are known to arise as near-horizon limits
of 1/2 supersymmetric $p$-brane solutions \cite{GT}. If the $D$-dimensional
supergravity is non-maximal then its $p$-brane solutions can usually be viewed
as intersections of branes of a maximal supergravity theory, \eg of
M-branes \cite{pap}. An example is the 1/2 supersymmetric self-dual
string solution of (1,1) D=6 supergravity, for which the near-horizon  geometry
is $adS_3\times S^3$ \cite{DGT}. This can be interpreted as the 1/4
supersymmetric $(1|M2,M5)$ string intersection of an M2-brane with an M5-brane ,
for which the near-horizon limit is a 1/2 supersymmetric $adS_3\times S^3\times
\bE^5$ vacuum \cite{boonstra}. 
 
The $(1|M2,M5)$ intersection is actually a special case of the 1/4 
supersymmetric $(1|M2,M5,M5)$ intersection in which the M2-brane intersects 
two M5-branes. The presence of the M2-brane is associated in this case with a
`generalized harmonic' function of the two sets of M5-brane coordinates \cite{hyper,loc}. {}For an
appropriate choice of this function there is again a non-singular horizon, near
which one finds a 1/2 supersymmetric $adS_3\times S^3\times S^3\times \bE^2$
vacuum of D=11 supergravity\footnote{This was originally found as an
$adS_3\times S^3\times S^3\times \bE^1$ solution of the heterotic string theory
\cite{ant} and was  shown in \cite{cow} to be the near-horizon limit of a
D=10 N=1 1/4 supersymmetric $(1|F1,NS5,NS5)$ intersection of a (fundamental)
string with two (Neveu-Schwarz) fivebranes.}. As argued in \cite{boonstrab}, and
confirmed here, the isometry supergroup of this D=11 vacuum solution is
\be
D(2|1,\alpha)_L\times D(2|1,\alpha)_R
\ee
where $\alpha$ is the ratio of the radii of the two 3-spheres. Each
$D(2|1,\alpha)$ factor contains an $Sl(2;\bR)\times SU(2)\times SU(2)$ bosonic
subgroup, with $\alpha$ being the relative weight of the two $SU(2)$ factors
(see e.g. \cite{toine}). In the limit in which one $S^3$ radius goes to infinity
we recover the $adS_3\times S^3\times \bE^5$ vacuum.

By adding momentum to the 1/4 supersymmetric $(1|M2,M5,M5)$ configuration
(along the string intersection) we arrive at the 1/8 supersymmetric
$(1|MW,M2,M5,M5)$ intersection. A curious feature of this case is that the
near-horizon limit is again the 1/2 supersymmetric $adS_3\times S^3\times
S^3\times \bE^2$ vacuum because a wave on $adS_3$ can be removed by a coordinate
transformation
\cite{HM}; this is simply a reflection of the fact that there are no
propagating gravitons in three-dimensions. The main purpose of this paper is to
present a new class of 1/8 supersymmetric $(1|MW,M2,M5,M5)$ solutions of D=11
supergravity with two independent rotation parameters. {}For an
appropriate choice of the functions parametrising this class of solution there
is a non-singular Killing horizon. The near-horizon limit is again the
$adS_3\times S^3\times S^3\times \bE^2$ vacuum. The isometry group is therefore
again $D(2|1,\alpha)_L\times D(2|1,\alpha)_R$. 

A number of supersymmetric rotating black hole and brane solutions have been
found previously. {}For example, there is a supersymmetric rotating black hole
solution of D=5 supergravity \cite{myers}, which can interpreted as an $S^1$
reduction of a rotating self-dual string solution \cite{myerstwo,larsen}.
It can also be interpreted as the dimensional reduction of a rotating
$(0|M2,M2,M2)$ intersection \cite{CY}. Many of these rotating brane
solutions are related by compactification and/or duality to some special case
of our rotating $(1|MW,M2,M5,M5)$ intersection. {}For example, the rotating
$(0|M2,M2,M2)$ intersection is dual to a rotating $(1|MW,M2,M5)$
intersection, which can then be dimensionally reduced to yield the rotating
self-dual string \cite{CY}. But the rotating $(1|MW,M2,M5)$ intersection is
just the special case of $(1|MW,M2,M5,M5)$ in which one M5-brane is omitted
and its associated rotation parameter set to zero. Thus, the new rotating brane
solution presented and analysed here constitutes a generalization of many
previous results on rotating branes.

\section{Rotating supersymmetric intersecting M-branes}

Consider two fivebranes and a membrane 
intersecting according to the pattern
\be
\matrix{
M5:&1&2&3&4&5& & & & & \cr
M5:&1& & & & & &7&8&9&10 \cr 
M2:&1& & & & &6& & & & \cr}.
\label{eq:snowy}
\ee
Our new 1/8 supersymmetric M-brane solution, which has
momentum flowing along the string intersection and carries angular momentum in
the relative transverse directions of the fivebranes ($\{2,3,4,5\}$ and $\{7,8,9,10\}$),
can be found within the following general class of solutions of D=11 supergravity. 
The bosonic sector of the D=11 supergravity Lagrangian is given by \cite{orig}
\be
\sqrt{-g}\left(R - {1\over 12}F^2\right)
+{2\over(72)^2}\epsilon^{M_1\cdots M_4N_1\cdots N_4P_1P_2P_3}
{}F_{M_1\cdots M_4}F_{N_1\cdots N_4}A_{P_1P_2P_3}
\ee
(in these conventions $d\star F+ F\wedge F=0$ 
and the equations of motion are solved by 
\bea
ds^2_{11} &=& g_1^{1/3}(g_2g_3)^{2/3}\big[ (g_1g_2g_3)^{-1}(-2dudv + g_4dv^2 +
2Adv) 
\label{newsol}\\
&&\qquad\quad + g_1^{-1}dz^2 + g_2^{-1}dx\cdot dx + g_3^{-1}dy\cdot dy\big]
\nn
{}F^{(4)} &=& {c_1\over2}\big\{du dv dg_1^{-1} - 
c_1c_2\star_x dg_2 - c_2\star_y dg_3 + dv\wedge A\wedge
dg_1^{-1} -g_1^{-1} dv\wedge dA\big\}\wedge dz
\nonumber
\eea
where the functions $g_1(x,y)$, $g_2(y)$, $g_3(x)$, $g_4(v,x,y,z)$ and
the 1-form
\be
A= A_i(x,y)dx^i + A_\alpha(x,y)dy^\alpha
\ee
will be specified below. The constants $c_1,c_2$ are signs and $\star_x$ and
$\star_y$ are the Hodge duals on the two Euclidean 4-spaces with
Cartesian coordinates $x^i$ and $y^\alpha$ (\eg $\star_x dg_2 =
(1/6)dx^idx^jdx^k\,\epsilon^{ijkl}\,\partial_lg_2$). 

The functions $g_2(y)$ and $g_3(x)$ are harmonic on their respective $\bE^4$
spaces and correspond to the two fivebranes. 
The function $g_1(x,y)$ corresponding to the membrane
is `generalized harmonic' \cite{hyper,loc}, \ie it satisfies
\be\label{geeone}
\big[g_3^{-1} \nabla_{(x)}^2 + g_2^{-1} \nabla_{(y)}^2\big] g_1 =0\ .
\ee
The function $g_4(v,x,y,z)$ corresponding to the gravitational wave 
is also a `generalized harmonic' function,
although in a slightly
more general sense than used hitherto. Specifically, it solves the equation
\be\label{avee}
\big[ g_3^{-1}\nabla_{(x)}^2 +
g_2^{-1}\nabla_{(y)}^2 +g_1 (g_2g_3)^{-1} \partial_z^2\big]g_4 =0
\ee
The field strength $F=dA$ of the 1-form $A$  satisfies
\bea
0&=&g_2^{-1}\pa_\a F_{\a j} +g_3^{-1}\pa_i F_{ij} +\pa_i g_3^{-1}[F_{ij} + 
{c_2\over 2} \epsilon_{ijkl} F_{kl}]\nn
 0&=&g_3^{-1}\pa_i F_{i \b} + g_2^{-1}\pa_a F_{\a\b} +
\pa_\a g_2^{-1}[F_{\a\b} + {c_1c_2\over 2} \epsilon_{\a\b\c\d} F_{\c\d}]
\label{feqn}
\eea

One may note the nesting of
the lower dimensional objects within the higher dimensional ones:
each fivebrane is localized within the worldvolume of the other one (except
along the string intersection), the membrane is localized within both of the
fivebrane worldvolumes, while the wave is localized within the worldvolumes of
both fivebranes and the membrane. Note that the dependence of the
wave-profile $g_4$ on the string direction $v$ is completely arbitrary.
This part of the solution is constructed following the
standard harmonic function rule \cite{pap}.
The new aspect of the solution, generalising \cite{tseytlinrotate}, 
comes from the introduction of the
1-form $A$. 
With an appropriate choice for the solution of
eq.~\p{feqn}, one can introduce angular momentum into both of the $\bE^4$
spaces. In order to maintain supersymmetry, this angular momentum
will involve simultaneous rotation in two orthogonal 2-planes within
each of these spaces.

Now we discuss the supersymmetries of this D=11 supergravity
solution. {}First we write the 11-metric in the form
\be
ds^2_{11} = -e^ue^u+e^v e^v + e^z e^z + e^i e^i + e^\alpha e^\alpha
\ee
with the orthonormal 1-forms
\bea
e^u &=& g_1^{-1/3}(g_2g_3)^{-1/6}g_4^{-1/2}\, (du- A)\nn
e^v &=& g_1^{-1/3}(g_2g_3)^{-1/6}g_4^{1/2}\,
              \left(dv -g_4^{-1}(du-A)\right)\nn
e^z &=& g_1^{-1/3}(g_2g_3)^{1/3}\, dz\nn
e^i &=& g_1^{1/6}g_2^{-1/6}g_3^{1/3}\, dx^i \nn 
e^\alpha &=& g_1^{1/6}g_2^{1/3}g_3^{-1/6}\, dy^\alpha
\eea
Let $\Gamma^A$ with $A=(u,v,z,i,\alpha)$ be the Dirac matrices in this basis.
The Killing spinor equations are then
\be
\bigg\{D + {1\over 144}
\big[e^A\Gamma_A{}^{BCDE} - 8e^B\Gamma^{CDE}\big]F_{BCDE}\bigg\}\epsilon =0
\ee
where $D$ is the Lorentz covariant exterior
derivative. The solutions are simply written as
\be
\epsilon = g_1^{-1/6}g_2^{-1/12}g_3^{-1/12}g_4^{-1/4}\epsilon_0
\label{killnot}
\ee
where $\epsilon_0$ is a constant spinor satisfying the constraints
\bea
\Gamma^{uv}\epsilon_0 &=& \epsilon_0\nn
\Gamma^z \epsilon_0 &=& c_1 \epsilon_0\nn
\Gamma_{(x)} \epsilon_0 &=& c_2 \epsilon_0\nn
\Gamma_{(y)} \epsilon_0 &=& c_1c_2 \epsilon_0 
\eea
where
\be
\Gamma_{(x)} = {1\over 24}\varepsilon_{ijkl}\Gamma^{ijkl}\qquad
\Gamma_{(y)} = {1\over 24}\varepsilon_{\alpha\beta\gamma\delta}
\Gamma^{\alpha\beta\gamma\delta}
\ee
The general solution \p{newsol}
still preserves 1/8 supersymmetry because these four conditions
are not all independent. This follows
since the 11-dimensional Dirac matrices satisfy
$
\Gamma^{uvz12341'2'3'4'}=1
$.

In order to fully specify a solution of D=11 supergravity, we have still to
choose the harmonic functions $g_2,g_3$, solve the `modified
harmonic' equations for $g_1$ and $g_4$, and solve (\ref{feqn}) for $A$.
We start by replacing the $x$-coordinates with polar coordinates
such that
\be
ds^2(\bE^4) = dr^2 + r^2d\Omega_3^2
\label{polar}
\ee
where $d\Omega_3^2$ is the 
metric on the 3-sphere.
A similar primed set of polar coordinates are chosen for the $y$-space.
Here, in order facilitate the analysis of the Killing spinors in the
next section, we choose coordinates on $S^3$ such that
\be
d\Omega_3^2 = {1\over4}(d\theta^2 + d\tilde\psi^2 + d\phi^2 + 2\cos\theta
d\tilde\psi d\phi))
\ee
with 
\be
0\le \theta < \pi\, ,\qquad 0\le \phi < 2\pi\, ,\qquad 
0\le \tpsi < 4\pi\, ,
\label{idee}
\ee
and similarly for the primed coordinates. This metric may be written in the
manifestly $SU(2)\times SU(2)$ invariant form 
\be
d\Omega_3^2 = {1\over4}(\tsigma_1^2 + \tsigma_2^2 + \tsigma_3^2)
\label{threesig}
\ee
where $\tsigma_i$ $(i=1,2,3)$ are the three left-invariant one-forms on
the group $SU(2)$: 
\bea
\tsigma_1 &=& -\sin\tpsi\, d\theta + \cos\tpsi\sin\theta\, d\phi \nn
\tsigma_2 &=& \cos\tpsi\, d\theta + \sin\tpsi\sin\theta\, d\phi \nn
\tsigma_3 &=& d\tpsi +\cos\theta\, d\phi\, .
\label{threesigma}
\eea

We now choose the harmonic functions corresponding to the fivebranes, $g_2$ and $g_3$,
to have single centres:
\be
g_3 = 1+ R^2/r^2 \qquad g_2 = 1+ {R'}^2/{r'}^2
\ee
where constants $R$ and $R'$ are related to fivebrane charges.
We will consider a solution of the remaining equations such
that the singularities at $r=0$, $r'=0$, and $r=r'=0$ are not genuine
curvature singularities but rather merely coordinate singularities.
To this end, we solve (\ref{geeone}) by setting
\be
g_1 =g_2g_3\, .
\ee
In addition we take 
\be
g_4=1 + p\, g_2 g_3
\ee
for some constant $p$. In solving (\ref{feqn}), we choose
\be
A = H'(y)B(x) + H(x)B'(y)
\ee
where $H$ and $H'$ are harmonic functions,
\be
H= 1 + {\ell^2\over r^2} \qquad H' = 1+ {{\ell'}^2\over {r'}^2}
\ee
and $B$ and $B'$ are 1-forms, on their respective $\bE^4$ spaces. In order
to solve (\ref{feqn}), the 2-form
field strengths, $G=dB$ and $G'=dB'$, must then satisfy
\be
G + \star_x G =0 \qquad G' + \star_y G' =0
\ee
where we have restricted to the case $c_1=c_2=1$, for convenience.
One solution of these anti-self-duality conditions is 
\be
B= {J\,\tsigma_3 \over 2\,r^2} \qquad B'= {J'\,\tsigma_3'\over 2\,{r'}^2}
\ee
for constants $J$ and $J'$, which can be shown (by examination of the
asymptotic metric as $r\rightarrow \infty$ and $r'\rightarrow \infty$) to
correspond to the magnitudes of angular momenta in the $x$ and $y$ 4-spaces. 
Thus as claimed, we have constructed a rotating 1/8 supersymmetric 
solution describing a $(1|MW,M2,M5,M5)$ intersection.

\section{Near-horizon limit}

{}For the remainder of the paper, we will be considering
a near-horizon limit of the preceding solution with both
$r/R,r'/R'<<1$. That is, a limit in which one approaches near both fivebranes
simultaneously. Note that the surface 
$r=r'=0$ is at infinite affine distance on spacelike
hypersurfaces (with finite $u$). One may construct this near-horizon
solution
with a scaling limit similar to that of \cite{juan}, and 
one obtains 
\bea
ds^2_{11} &=& (RR')^{-2}\left[ -2(rr')^2 du + \hat J
\tsigma_3  + \hat J' \tsigma_3' \right]dv + pdv^2  \nn
&& + R^2{dr^2\over r^2} + 
{R'}^2 {dr'^2\over r'^2} + R^2d\Omega_3^2 + 
{R'}^2d(\Omega'_3)^2 + dz^2
\eea
where
$\hat J = \ell'^2 J$ and  $\hat J' = {\ell}^2 J'$.
In terms of the new coordinates
\bea
\rho &=& rr' \nn
z' &=& {R^2\over {\sqrt{R^2 + {R'}^2}}}\log r -  {{R'}^2\over {\sqrt{R^2 + {R'}^2}}}
\log r'\nn
w &=& {v\over R^2 {R'}^2}\nn
\psi&=&\tpsi+{2\hat J\over  R^2}w\nn
\psi'&=&\tpsi'+{2\hat J'\over  {R'}^2}w\ ,
\label{coordtrans}
\eea
the 11-metric becomes
\be
ds^2_{11} = -2\rho^2 dudw + \gamma^2{d\rho^2\over \rho^2} + Q^2dw^2
+ R^2 d\Omega_3^2 + 
{R'}^2d{\Omega'}^2_3+dz^2+{dz'}^2\ .
\label{nearsol}
\ee
with
\be
Q^2= \left(pR^4{R'}^4-{\hat J^2\over  R^2}-{{\hat J}'^2\over  {R'}^2}\right)
\label{cue}
\ee
and
\be
\gamma = {R R' \over {\sqrt{R^2 + {R'}^2}}}\, .
\ee
This metric has the simple form of $adS_3\times S^3\times S^3\times \bE^2$
with $\Lambda_{adS_3}=-{\gamma^{-2}}$. In these coordinates, the 
near-horizon four-form field strength is given by
\be
{}F=dz\wedge({Vol(S^3_R)\over R} + {Vol(S^3_{R'})\over R'}
-{Vol(adS_3)\over \gamma})
\label{nearfield}
\ee
where $Vol(X)$ denotes the volume form on the corresponding
three-dimensional space $X$.

Note that up to this point we have not considered any of the coordinates to be
periodically identified, other than the 3-sphere identifications \p{idee}. Thus
there are no global subtleties in implementing the above coordinate
transformation
\p{coordtrans}. In particular, one finds that the remnant of the angular momentum
in this near-horizon limit is  a coordinate artifact because the transition from
$\tpsi,\tpsi'$ to $\psi,\psi'$ removes all mixing of $v$ with the angular
coordinates. 
This generalizes a similar observation made recently in the context
of the near-horizon limit of certain spinning five-dimensional black holes
\cite{larsen}.
There is still a remnant of the gravitational wave in the near-horizon 
metric in the form of the $dw^2$ term. This can also be removed by a coordinate
transformation \cite{HM} but we choose not to do this here because
compactifications will typically require periodic identification of 
$w$, in which case the coordinate transformation that removes the $dw^2$
term could only be implemented locally. In general, the angular momenta and
gravitational wave of the full solution will leave their imprint in the
near-horizon solution in the form of the global identifications which complete
the specification of the compactified geometry.

\section{Near-horizon Killing spinors}

Now we wish to determine the supersymmetry enhancement which
arises in the near-horizon limit above by explicitly solving for the
Killing spinors. Recall that the Killing spinors satisfy 
\be
\left\lbrace D+{1\over144}\left[e^A\Gamma_A{}^{BCDE}-8e^B\Gamma^{CDE}\right]
{}F_{BCDE}\right\rbrace\epsilon=0
\label{killeq}
\ee
where the covariant exterior derivative is given by
\be
D=d+{1\over4}e^A\omega_{ABC}\Gamma^{BC}\equiv
d+e^AM_A\ .
\label{covext}
\ee

A suitable choice of frame 1-forms for the near-horizon 11-metric \p{nearsol} is
\bea
e^u &=& {\rho^2\over Q} du \nn
e^w &=& Q\,dw - {\rho^2\over Q} du\nn
e^\rho &=& \g{d\rho\over \rho}\nn
e^i &=& {R\over 2}\, \sigma_i \qquad (i=1,2,3)\nn
e^\alpha &=& {R'\over 2}\, \sigma'_\alpha \qquad (\alpha=1',2',3')\nn
e^z&=&dz \qquad\qquad e^{z'}=dz'
\label{frame}
\eea
where the forms $\sigma_{i}$ are defined as in \p{threesigma}
with $\theta,\phi,\psi$, and similarly for the primed angles.
The matrices $M_A$ appearing in the covariant
derivative \p{covext} are
\bea
M_u &=& -{1\over 2\gamma} \Gamma^{w\rho}-{1\over \gamma}\Gamma^{u\rho} \nn
M_w &=& -{1\over2\gamma}\Gamma^{u\rho} \nn
M_\rho &=& {1\over2\gamma} \Gamma^{wu}\nn
M_1 &=& {1\over2R}\Gamma^{23} \qquad ({\rm and\ cyclic})\nn
M_{1'} &=& {1\over2R'}\Gamma^{2'3'} \qquad ({\rm and \ cyclic})
\eea
while $M_z=0=M_{z'}$.

We now define matrices $N_A$  by
\be
e^AN_A\equiv
{1\over 144}\big(e^A\Gamma_A{}^{BCDE} -8e^B\Gamma^{CDE}\big)F_{BCDE}\ ,
\ee
so that the Killing spinor equations \p{killeq} become 
$(\partial_\mu + e^A{}_\mu(M_A+N_A))\epsilon=0$.
In particular, using \p{nearfield} one has 
\be
N_z = -{1\over3}\left[{1\over R}\Gamma^{123} + {1\over R'}\Gamma^{1'2'3'} 
-{1\over \gamma}\Gamma^{uw\rho}\right]
\ee
Therefore the Killing spinors $\epsilon$ will be independent of $z$
provided that $N_z\epsilon=0$, which is equivalent to
\be
\Gamma\epsilon = \epsilon
\label{constraint}
\ee
where
\be
\Gamma= {\gamma\over R}\Gamma^{uw\rho 123} + 
{\gamma\over R'}\Gamma^{uw\rho 1'2'3'}\ .
\label{defcon}
\ee
Note that $\Gamma^2=1$. One also finds that $N_{z'}=(1/2)\Gamma^{zz'}
N_z$. 
Therefore the constraint \p{constraint} also guarantees that the
Killing spinors are independent of $z'$.
The remaining $N_A$ are
\bea
N_u &=& \left({1\over6\gamma}\Gamma^{\rho w}\Gamma + {1\over3\g}\Gamma^{\rho w}
\right)\Gamma^z\nn
N_w &=& \left({1\over6\g}\Gamma^{u\rho}\Gamma + {1\over3\g}\Gamma^{u\rho} 
\right)\Gamma^z\nn
N_\rho &=& \left({1\over6\g}\Gamma^{wu}\Gamma + {1\over3\g}\Gamma^{wu}
\right)\Gamma^z \nn
N_1 &=& \left({1\over6\g}\Gamma^{1w\rho u}\big[1-\Gamma\big] +
{1\over2R}\Gamma^{23}\right)\Gamma^z\qquad ({\rm and\ cyclic})\nn
N_{1'} &=& \left({1\over6\g}\Gamma^{1'w\rho u}\big[1-\Gamma\big] +
{1\over2R'}\Gamma^{2'3'}\right)\Gamma^z\qquad ({\rm and\ cyclic})
\eea

On the $\Gamma=1$
eigenspace the remaining Killing spinor equations now become
\bea
0 &=& \big(\partial_u - {\rho^2\over Q\g}(\Gamma^{w\rho}-\Gamma^{\rho u})
\Gamma_+\big)\epsilon \nn
0 &=& (\partial_w - {Q\over \g}\Gamma^{u\rho}\Gamma_- )\epsilon \nn
0 &=& (\partial_\rho + \rho^{-1}\Gamma^{wu}\Gamma_+)\epsilon \nn
0 &=& (\partial_\theta + {1\over2}(\sin\psi\,\Gamma^{32} +
\cos\psi\,\Gamma^{31})\Gamma_+)\epsilon \nn
0 &=& (\partial_\phi + {1\over2}(\cos\psi\sin\theta\,\Gamma^{23} 
+\sin\psi\sin\theta\,\Gamma^{31} + \cos\theta
\Gamma^{12})\Gamma_+)\epsilon \nn
0 &=& (\partial_\psi +{1\over2}\Gamma^{12}\Gamma_+)\epsilon 
\label{lotseq}
\eea
and similarly for the equations in the primed angles,
where we have defined the projection operators
\be
\Gamma_\pm={1\over2}\left(1\pm\Gamma^z\right)\ .
\label{project}
\ee

The solution can be
expressed in terms of a constant spinor $\chi$ subject to the single
constraint \p{constraint}
\be
\Gamma\chi = \chi
\label{constraintb}
\ee
As $\Gamma$ commutes with both $\Gamma^z$ and $\Gamma^{wu}$ this
condition is compatible with the decomposition
\be
\chi = (\chi_+^+ +\chi_+^-) + (\chi_-^+ + \chi_-^-)
\ee
where
\be
\Gamma^z \chi_\pm = \pm \chi_\pm \qquad \Gamma^{wu}\chi^\pm = \pm \chi^\pm\ .
\label{projects}
\ee

Since $\Gamma_+$ annihilates $\chi_-$, the corresponding Killing spinors
are independent of all coordinates except $w$. The $w$ equation is solved
by
\be
\epsilon_- = e^{{Qw\over \g}\Gamma^{u\rho}}(\chi_-^+ + \chi_-^-)
\label{resulta}
\ee
Similarly $\Gamma_-$ annihilates $\chi_+$, so the corresponding
Killing spinors are independent of $w$. The remaining equations \p{lotseq}
are solved by
\be
\epsilon_+  = ({1\over\rho} -{2u\rho\over Q\g}\Gamma^{\rho u})\Omega\Omega'
\chi_+^+ +\rho\,\Omega\Omega'\chi_+^-
\label{resultb}
\ee
where 
\be
\Omega(\theta,\phi,\psi)=e^{{1\over2}\Gamma^{21}\psi}
e^{{1\over2}\Gamma^{13}\theta}e^{{1\over2}\Gamma^{21}\phi}
\label{OOmega}
\ee
and similarly for $\Omega'(\theta',\phi',\psi')$.

Hence given the single constraint \p{constraint}, we conclude that
the near-horizon solution, \p{nearsol} and \p{nearfield}, preserves
one-half the supersymmetries. This represents a four-fold increase
over that for the full solution presented in the first section.
Of course this counting
assumes that only the angular coordinates have periodic identifications.
With a reduction on $w$, (naively at least) the $w$ dependence of
$\epsilon_-$
rules these out as admissible supersymmetries, and the near-horizon
solution would only preserve one-quarter the supersymmetries,
as expected. We will discuss this further in section 7.

Recall that the determination of Killing spinors in section 2
was done for the general solution \p{newsol} without a particular
choice of metric functions. {}For the purpose of comparing with
the present results it is convenient to repeat this analysis in
polar coordinates \p{polar}. With the assumption that the only 
$x^i$ and $y^\alpha$ dependence is on the respective radial coordinates,
$r$ and $r'$, \ie $g_1=g_1(r,r'),$ $g_2=g_2(r')$, $g_3=g_3(r)$,
$g_4=g_4(r,r',z)$, one
finds \p{killnot} is replaced by
\be
\epsilon = g_1^{-1/6}g_2^{-1/12}g_3^{-1/12}g_4^{-1/4}\Omega\Omega'\epsilon_0\ .
\label{killnotb}
\ee
Hence if we look at the near-horizon limit of these spinors,
 we find
\be
\epsilon = r r'\,\Omega\Omega'\epsilon_0=
\rho\,\Omega\Omega'\epsilon_0\ .
\ee
So as expected, we see that these Killing spinors match up
to the  $\epsilon^-_+$ spinors in \p{resultb}. That is, 
only $\epsilon^-_+$ supersymmetries may be extended into the asymptotic
regions of the solution, \eg large $\rho$. 

\section{Killing spinors and the isometry superalgebra}

Any supergravity solution can be presented as a configuration of tensors on
superspace. The local isometry supergroup of the solution is, by definition,
the supergroup generated by the Killing vector superfields, i.e.
those 
vector superfields which leave the superspace
configuration invariant (see \cite{dewitt} for details). 
It might seem that a determination of the isometry superalgebra of a given
supergravity solution would require that one first find its superspace
presentation, which is a very laborious task. {}Fortunately, a shortcut is
possible, at least for solutions that are purely bosonic.  The method, which we
explain below, relies on the fact that any pair of Killing spinors determines a
Killing vector field. We shall now explain this point.

We begin by noting  that given two D=11 Killing spinors, $\zeta$ and $\zeta'$,
then the D=11 vector field
\be
v=\bar\zeta\Gamma^M\zeta'\, \partial_M
\ee
is Killing. The proof is as follows: We first observe that
\be
\overline{\Gamma_{(n)}\zeta} = (-1)^{{n(n+1)\over 2}}\bar\zeta \Gamma_{(n)} 
\ee
where $\Gamma_{(n)}$ is the antisymmetrized product of $n$ of the Dirac matrices
$\Gamma^A$. It follows, for any spinor $\zeta$, that
\be\label{kilone}
{\overline {D_M\zeta}} = \partial_M\bar\zeta
-{1\over4}\bar\zeta\,\Gamma^{AB}\, \omega_{MAB}
=D_M\overline{\zeta}\, ,
\ee
and that a Killing spinor $\zeta$ satisfies
\be
\overline{D_M\zeta} = {1\over 144} \bar\zeta \big[ \Gamma_M{}^{NPQR} +
8\delta_M{}^N\Gamma^{PQR}\big] F_{NPQR}
\ee
It further follows from (\ref{kilone}) that
\be
D_M v_S ={\overline {D_M\zeta}}\,\Gamma_S \zeta' + \bar\zeta \Gamma_S D_M \zeta'
\ee
Using the Killing spinor condition in both terms on the
right hand side we deduce that
\be
D_M v_S = {1\over 144} \bar\zeta\bigg[ 2{\Gamma_{M}{}^{NPQR}}_S 
+48\, \delta_M{}^N \delta_S{}^P\Gamma^{QR}\bigg]\zeta'F_{NPQR}
\ee
and hence that
\be
D_{(M} v_{S)} =0\, 
\ee
as required for a Killing vector field. 

We now turn to the explanation how the above observations can
be used to determine the isometry superalgebra of a given supergravity
solution (and hence its local isometry supergroup) from a knowledge 
of its Killing spinors. An expansion of the D=11 supergravity action about 
the solution of interest yields currents $(T^{mn},j^m_\alpha, K^{mnp})$ with
background covariant conservation conditions determined by the gauge
transformations of the fluctuation fields. These currents must belong to a
supermultiplet with respect to the supersymmetries of the background associated
with Killing spinors. This statement can be formalized in terms of the
once-integrated current algebra anticommutator
\be\label{currental}
\{Q_F(\zeta),j^m\} = {1\over2} T^{mn}\Gamma_n\zeta +
{1\over2}K^{mnp}\Gamma_{np}\zeta
\ee
where
\be
Q_F(\zeta) = \int_\Sigma dS_m\, \bar\zeta j^m
\ee
is the ({}Fermionic) charge of the fluctuation fields associated with the
background Killing spinor $\zeta$. {}For fluctuation fields that fall off
sufficiently rapidly towards the boundary at infinity on the spacelike
hypersurface $\Sigma$, these charges will be time-independent by virtue of the
conservation condition on $j^m$ and the Killing spinor condition obeyed by
$\zeta$. 

Integrating the relation (\ref{currental}) and discarding the integral over
$K^{mnp}$ on the grounds that it could be non-zero only in the presence of a
membrane source, we find that
\be
\{Q_F(\zeta),Q_F(\zeta')\} = Q_B(v=\zeta'\Gamma\zeta)
\ee
where
\be
Q_B(v) = \int_\Sigma dS_m\, v^nT^m{}_n\, .
\ee
Since $v$ is Killing this (Bosonic) charge is also time-independent.
What this shows is that the determination of the linear combination of 
Killing vector fields 
associated with any pair of Killing spinors is equivalent to the determination
of the linear combination of bosonic charges in the isometry superalgebra that
appear in the anticommutator of any pair of fermionic charges in this algebra.

We shall now use this method to determine the isometry superalgebra of the
near-horizon limit of our new rotating brane solution and of some of its
special cases.
Of course, there may be additional bosonic isometries that are
not found in the above way. In this case the full isometry superalgebra will be
the direct product of the superalgebra as determined by the above method with a
purely bosonic algebra.

\section{The isometry supergroup of $AdS_3\times S^3\times S^3$}

Using the arguments of the last section we conclude that
to determine the isometry supergroup of the near horizon geometry
of our rotating brane solution, which is
simply
$AdS_3\times S^3\times S^3$, we need to construct the Killing vectors
associated to the Killing spinors \p{resulta} and \p{resultb}.
We will show that the answer is given by $D(2|1,\alpha)\times D(2|1,\alpha)$

To begin, we first observe that for any two Killing spinors $\epsilon,\e'$ we
have 
\be
\bar\e\Gamma^z\e'=\bar\e\Gamma^{z'}\e'=0
\ee
since they lie on the $\Gamma=1$ eigenspace. In addition we have
\be
\bar\epsilon_+\Gamma^A\epsilon_- =0
\ee
for all $A$ (using, for example, $\{\Gamma^A,\Gamma^z\}=0$ for $A\ne z$).  
It follows that the only Killing vectors obtainable from Killing spinors are
\be
v_{++} = (v_{++}^{++},v_{++}^{+-},v_{++}^{--})
\quad{\rm and}\quad
v_{--} =  (v_{--}^{++},v_{--}^{+-},v_{--}^{--})
\ee
where, \eg  $v_{--}^{+-} = \bar\epsilon_-^+ \Gamma^A \epsilon_-^- \, \tilde
e_A$. \footnote{Note that our notation here means, for example,
$\epsilon^-_+$ is the Killing spinor depending on $\chi^-_+$.
However, unlike $\chi^-_+$, $\epsilon^-_+$ does not have
a simple projection under $\Gamma^{wu}$, \ie
$\Gamma^{wu}\epsilon_-^+\ne+\epsilon_-^+$.}
Here, we are using the dual basis vectors $\tilde e_A$
\bea\label{dualvf}
\tilde e_u &=& Q\rho^{-2}\partial_u +Q^{-1}\partial_w\nn
\tilde e_w &=& Q^{-1}\partial_w  \nn
\tilde e_\rho &=&{\rho\over \gamma} \partial_\rho \nn
\tilde e_i &=& {2\over R}\xi_i^R \nn
\tilde e_\alpha &=& {2\over R'}\xi_\alpha^R\nn
\tilde e_z &=& \partial_z\qquad\qquad
\tilde e_{z'}=\partial_{z'}
\eea
where $\xi^R$ are left-invariant vector fields on each $S^3$ dual 
(that generate right actions) which are dual 
to the left-invariant
one-forms $\sigma$ and whose explicit form is
\bea
\xi_1^R &=& -\sin\psi\partial_\theta + \cos\psi\cosec\theta\partial_\phi
-\cot\theta\cos\psi\partial_\psi \nn
\xi_2^R&=& \cos\psi\partial_\theta + \sin\psi\cosec\theta\partial_\phi
-\cot\theta\sin\psi\partial_\psi \nn
\xi_3^R &=& \partial_\psi \, .
\eea
Later the right-invariant vector fields $\xi^L$ (that generate left actions)
will also appear. They
are given by 
\bea\label{sighleft}
\xi^L_1 &=& \sin\phi\partial_\theta + \cot\theta\cos\phi \partial_\phi
-\cos\phi\cosec\theta \partial_\psi\nn
\xi^L_2 &=& \cos\phi\partial_\theta - \cot\theta\sin\phi\partial_\phi
+ \sin\phi\cosec\theta \partial_\psi\nn
\xi^L_3 &=& \partial_\phi\, .
\eea
The commutation relations are given by 
\be\label{comrel}
{[\xi_i^R,{\xi_{j}^{R}}]}=-\epsilon_{ijk}\xi_k^R
\qquad
{[{\xi_{i}^{L},\xi_{j}^{L}}]}=\epsilon_{ijk}\xi_k^L
\qquad
{[{\xi_i^R,\xi_j^L}]}=0
\ee

There are a number of useful identities in the computation of
the Killing vectors. {}First, we observe the fact that
\be
\bar\chi^\pm\Gamma^\rho\chi^\pm =
\bar\chi^\pm\Gamma^i\chi^\pm = \bar\chi^\pm\Gamma^\alpha\chi^\pm =0
\ee
because $\Gamma^\rho,\Gamma^i$ and $\Gamma^\a$ all commute with $\Gamma^{wu}$. 
Also
\be
\bar\chi^\pm\Gamma^u\chi^\mp =\bar\chi^\pm\Gamma^w\chi^\mp =0
\ee
because $\Gamma^u$ and $\Gamma^w$ anticommute with $\Gamma^{wu}$. Similarly
\be
\bar\chi^\pm\Gamma^{w\rho i}\chi^\mp =0 \qquad
\bar\chi^\pm\Gamma^{w\rho \alpha}\chi^\mp =0\ .
\ee
{}Finally note from the definition of the superscript indices on the
constant spinors that $\Gamma^w\chi^\pm=\pm\Gamma^u\chi^\pm$.

Now a calculation  yields
\bea
v_{--}^{++} &=& {2\over\g}(\bar\chi_-^+\Gamma^u \chi_-^+)\ell_+^R\nn
v_{--}^{+-} &=& {2\over\g}(\bar\chi_-^+\Gamma^\rho\chi_-^-)\ell_0^R
+ {2\over R} 
(\bar\chi_-^+\Gamma^i\chi_-^-)\xi_i^R + {2\over R'}
(\bar\chi_-^+\Gamma^\alpha\chi_-^-)\xi_\alpha^R\nn
v_{--}^{--} &=& {2\over\g}(\bar\chi_-^-\Gamma^u \chi_-^-)\ell_-^R
\label{right}
\eea
where 
\bea
\ell_\pm^R &=&  {\g Q\over2\rho^2}\cosh{2Qw\over \g}\partial_u
         +  {\g\over 2Q}(\cosh{2Qw\over \g} \pm1) \partial_w
         -{\rho\over 2} \sinh{2Qw\over \g}\partial_\rho \nn
\ell_0^R &=& -{\g Q\over2\rho^2}\sinh{2Qw\over \g}\partial_u
         -  {\g\over 2Q}\sinh{2Qw\over \g} \partial_w
         +{\rho\over 2} \cosh{2Qw\over \g}\partial_\rho)
\label{rightv}
\eea
These obey the $SO(2,1)$ commutation relations
\be
[\ell_+^R,\ell_-^R] =-2\ell_0^R \qquad [\ell_0^R,\ell_\pm^R]=\pm\ell_\pm^R
\label{soads}
\ee

The $v_{++}$ Killing vectors are
\bea
v_{++}^{++} &=& {2\over\g}(\bar\chi_+^+\Gamma^u\chi_+^+) \ell_+^L \nn
v_{++}^{+-} &=&  {2\over \g}(\bar\chi_+^+\Gamma^\rho\chi_+^-) \ell_0^L
+ {2\over R}(\bar\chi_+^+\Gamma^i\chi_+^-)\xi_i^L
+ {2\over R'} (\bar\chi_+^+\Gamma^\alpha\chi_+^-)\xi_\alpha^L\nn
v_{++}^{--} &=& {2\over \g}(\bar\chi_+^-\Gamma^u\chi_+^-) \ell_-^L
\label{left}
\eea
where
\bea
\ell_+^L&=&({\g Q\over 2\rho^4}+{2u^2\over Q\g})\partial_u+{\g\over Q\rho^2}
\partial_w-{2u\rho\over Q\g}\partial_\rho\nn
\ell_-^L &=& {\g Q\over 2}\partial_u \nn
\ell_0^L &=& {\rho\over 2}\partial_\rho -u\partial_u
\label{leftv}
\eea
which satisfy the standard $SO(2,1)$ commutation relations
\be
[\ell_+^L,\ell_-^L] =2\ell_0^L \qquad [\ell_0^L,\ell_\pm^L]=\mp\ell_\pm^L
\label{soadsl}
\ee
and in addition commute with the $R$ generators.
In deriving \p{left} we have used the fact that
\be
\Omega^{-1}\Gamma^i \Omega \equiv \Gamma^j {R_j}^i(\Omega)
\ee
with 
\be
{R_j}^i(\Omega)\xi^R_i =\xi^L_j
\ee

{}From \p{right} and \p{left}
we conclude that the near-horizon isometry supergroup 
is given by
\be\label{isom}
D(2|1,\alpha)_L\times D(2|1,\alpha)_R
\ee
where
\be
\alpha = {R'\over R}
\ee
which is the ratio of the radii of the two 3-spheres.

\section{Killing spinors and reduction}

One might consider constructing solutions of lower-dimensional
supergravity from D=11 supergravity solutions when the latter have Killing
symmetries. Here, we wish to consider the effect of such dimensional
reduction on the Killing spinors. Naively, one expects that the only
Killing spinors to survive will be those which are independent of the
internal coordinates on which one is reducing.
More precisely, when reducing on a Killing vector $k$, 
we must require that the Killing spinors have a vanishing
Lie derivative under $k$. The Lie derivative of a spinor $\epsilon$ with respect
to an arbitrary vector field is ill-defined, but with respect to a Killing vector
field it is given by (see, \eg \cite{hyper})
\be
{\cal L}_k\epsilon=i_kD\epsilon +{1\over8}\Gamma^{mn}(d\, k)_{mn}\epsilon
\ee
where $D$ is the covariant derivative defined in \p{covext},
and $(d\, k)$ is the exterior derivative of the 1-form $k_m dx^m$ dual to
$k^m\partial_m$. Killing spinors satisfy the supersymmetry Killing
equations: $(D_m+N_m)\epsilon=0$. Thus, a vanishing Lie derivative of
$\epsilon$ implies
\bea
0&=& k^m D_m\epsilon+{1\over8}\Gamma^{mn}(d\, k)_{mn}
\epsilon\nn
&=&\left({1\over8}\Gamma^{mn}(d\, k)_{mn}-k^mN_m\right)\epsilon 
\equiv P\epsilon\ .\label{newcons}
\eea
In other words, the vanishing Lie derivative condition reduces to a simple
algebraic $\Gamma$-matrix constraint ($P\epsilon=0$) on the Killing spinors.
Dimensional reduction will therefore reduce the number of Killing spinors to
those satisfying this constraint. The details will, of course, depend on the
solution and the particular choice of $k$, so we shall illustrate the procedure
with a number of simple examples.

\subsection{Zero Angular Momentum}
The non-rotating
near-horizon metric \p{nearsol} may be written as
\bea
ds^2_{11} &=&-\left({\rho^2\over Q}du\right)^2+\left({Q}dw
-{\rho^2\over Q}du\right)^2+\g^2\left({d\rho\over\rho}\right)^2\nn
 &&\qquad+{R^2\over 4}(\sigma_1^2+\sigma_2^2+\sigma_3^2)
 +{{R'}^2\over 4}(\sigma_1'{}^2+\sigma_2'{}^2+\sigma_3'{}^2)+dz^2+dz'^2
\label{again}
\eea
There are
a number of simple Killing vectors upon which we will consider
reducing the solution: $\partial_z$, $\partial_{z'}$, $\partial_w,$
$\partial_\psi$ and $\partial_{\psi'}$. Note that 
demanding that (any of the first three of)
these Killing vectors have closed orbits, will imply 
global identifications on 
the $adS_3\times S^3\times S^3\times \bE^2$ geometry.

The reduction on $\partial_z$ or $\partial_{z'}$ is trivial. 
In this
case, $d\, k=0$ and so the constraint \p{newcons} reduces to
$P\epsilon=-N_{z,z'}\epsilon=0\ .$
However, this is equivalent to
the constraint \p{constraint} already imposed on all of the Killing
spinors. Of course, \p{constraint} was derived from requiring the
Killing spinors be independent of $z$ and $z'$. Hence it is no
surprise that all of the Killing spinors survive unchanged when the
theory is reduced on these two directions, \ie there is no
reduction in the number of supersymmetries.

Next consider a reduction on $\partial_w$. 
{}First let us note that the metric has precisely
the one that one would adopt for a standard Kaluza-Klein compactification
for a reduction on $w$. The reduced metric would correspond to \p{again}
without the $(e^w)^2$ term, \ie
\be
ds^2 =-\left({\rho^2\over Q}du\right)^2+\g^2\left({d\rho\over\rho}\right)^2
+\ldots
\label{adstwo}
\ee
and the off-diagonal component $g_{wu}$ would become a gauge field
$-{\rho^2/ Q}\,du$ in the
lower-dimensional theory. As is apparent in \p{adstwo},
$adS_3$ is replaced by $adS_2$ in the reduced geometry.
If the frame \p{frame} (without
$e^w$) is chosen to describe the reduced geometry, the form of the
Killing spinors will be unchanged up to the
additional constraint \p{newcons}.
In this case, we have $k=Q\te_w$, and as a 1-form, $k
=Qe^w$. Hence one finds $d\,k={2Q\over \g} e^u e^\rho$, and 
from \p{newcons}
$P={Q\over \gamma}\Gamma^{u\rho}\,\Gamma_-$. Hence the Killing spinors surviving the
reduction must satisfy
$\Gamma_-\epsilon=0$
which picks out the $\epsilon_+$ spinors in \p{resultb}. This result
then agrees with the naive expectation that one should chose the
spinors independent of $w$.

Reducing on $\partial_\psi$ also produces
an interesting lower-dimensional solution. (Of course, a reduction on
$\partial_{\psi'}$ completely parallels the following analysis.)
Recalling the definition \p{threesigma} for the $\sigma_i$'s,
we note that the metric \p{again} is again adapted for a Kaluza-Klein
reduction on $\psi$. In this case, the reduced metric becomes
\bea
ds^2&=&{R^2\over4}(\sigma_1^2+\sigma_2^2)+\ldots\nn
&=&{R^2\over4}(d\theta^2+\sin^2\theta\,d\phi^2) +\ldots
\label{sphtwo}
\eea
and so one of the $S^3$ factors is replaced by $S^2$ in the reduced
geometry. The latter also carries a monopole gauge field arising
from $g_{\psi\phi}$. Here, we have $k=(R/2)\te_3$,
and as a 1-form, $k=(R/2)e^3$. Then \p{newcons} yields
$P={1\over 2}\Gamma^{12}\,\Gamma_-$, and so the Killing spinors surviving
this reduction again satisfy
$\Gamma_-\epsilon=0$.
Thus the $\epsilon_+$ spinors correspond to supersymmetries in the
reduced geometry.

At first sight, this is a surprise since it
is the $\epsilon_-$ spinors \p{resulta} which are independent of $\psi$.
The resolution of this puzzle comes from realizing that $e^1$ and $e^2$
can not be used as orthonormal 1-forms in the reduced theory, as
they are $\psi$ dependent --- see \p{threesigma}. Rather they should be
replaced by the 1-forms, 
\eg
\be
\hat e^1 ={R\over2} \sin\theta\,d\phi \qquad \hat e^2={R\over2}d\theta\ .
\label{newframe}
\ee
In this case, the two sets of 1-forms are related by a simple rotation
acting in the 1-2 subspace
\be
\hat e^a=L^a{}_b\,e^b\quad{\rm with}\quad
L=\pmatrix{\cos\psi&\sin\psi\cr
           -\sin\psi&\cos\psi\cr}
 =\exp\left[\pmatrix{0&1\cr -1&0\cr}\psi\right]
\label{rotate}
\ee
If Lorentz vectors are rotated by
$L^a{}_b=\exp(\omega^a{}_b)$,
then the corresponding transformation of spinors is
\be
\widetilde{L}=\exp\left({1\over4}\omega_{ab}\Gamma^{ab}\right)\ .
\label{rotatespin}
\ee
Specifically for \p{rotate}, we have
\be
\widetilde{L}_\psi=\exp\left(-{1\over2}\Gamma^{21}\psi\right)\ .
\label{rotatepsi}
\ee
Hence reducing on $\partial_\psi$ requires modifying the frame, and
 in doing so the precise form of the Killing spinors changes by
\be
\hat\epsilon=\widetilde{L}_\psi\epsilon=e^{-{1\over2}\Gamma^{21}\psi}
\epsilon\ .
\ee
However, this transformation precisely cancels the $\psi$ dependence
of the $\epsilon_+$ spinors, and thus the $\hat\epsilon_+$ appear
as Killing spinors in the reduced theory. Here, we should note that
since $\Gamma_-$ commutes with $\widetilde{L}_\psi$, the form of
the constraint $\Gamma_-\epsilon=0$ is identical for both $\epsilon$ and
$\hat\epsilon$.

Note that in both of the latter two reductions, the constraint
\p{newcons} reduces the number of supersymmetries by $1/2$.
{}Furthermore in selecting out the
$\epsilon_+$ spinors, the reduced supersymmetries include those,
\ie $\epsilon_+^-$,
that can be extended into the asymptotic regions of the full solution.

Having obtained the Killing spinors in the reduced solution
we can determine the corresponding superalgebras
by following the
steps in section five and six. In both of the above cases, the Killing spinors
have the form $\hat \epsilon_+$. 
The Killing vectors are then obtained
by determining
\be
\bar {\hat \epsilon}_+ \Gamma^a {\hat \epsilon}_+ \tilde{\hat e}_a
\ee
where the sum is now over all indices excluding the coordinate
that one reduces on and the $ \tilde{\hat e}_a$ are the dual vector fields
in the reduced spacetime.

Let us first consider the reduction on $\partial_w$ to give $adS_2\times S^3\times
S^3$. Although the frame \p{frame}
without $e_w$ is a suitable frame for the reduced spacetime
the dual vector fields \p{dualvf} are not: instead we must now use 
$\tilde{\hat e}_u =Q\rho^{-2}\pa_u$.
Taking this into account we find the Killing vectors as in \p{left} 
with the only difference being that we drop $\partial_w$ from $\ell^L_+$ in 
\p{leftv}. This means that the superalgebra contains
a factor $D(2|1,\alpha)$. Combining this with the
bosonic symmetries
that don't arise from Killing spinors, we conclude that
the symmetry algebra is
given by 
\be
\label{fig}
D(2|1,\alpha)\times SU(2)\times SU(2).
\ee

Next consider the reduction on $\partial_\psi$ to obtain $adS_3\times S^2\times
S^3$. We again obtain \p{left} 
but now with the Killing vectors $\xi_L$ replaced by Killing vectors obtained by setting
$\psi=0$ and dropping $\partial_\psi$ terms in  the expressions for $\xi_L$
in \p{sighleft}. 
In this case the symmetry algebra
is $D(2|1,\alpha)\times SO(2,1)\times SU(2)$.

\subsection{Adding Angular momentum}

Recall that the remnants of the angular momenta were eliminated
by the coordinate transformation \p{coordtrans} in the near-horizon limit.
After reduction on $w$, $\psi$ or $\psi'$, such a transformation
would not be allowed and so we should reconsider these reductions
in the presence of the angular momentum. {}First, we
must insert the angular momenta back in the metric \p{again}. This is easily
done as we simply undo part of the original coordinate transformation
\p{coordtrans},
reintroducing $\tpsi,\tpsi'$
\bea
\psi&=&\tpsi+{2\hat J\over R^2}\,\tw\nn
\psi'&=&\tpsi'+{2\hat J'\over {R'}^2}\,\tw\nn
w&=&\tw
\label{oldcoord}
\eea
which in our angular forms yields
\bea
\sigma_3&=&\tsigma_3+{2\hat J\over R^2}\,d\tw\nn
\sigma_3'&=&{\tsigma}'_3+{2\hat J'\over {R'}^2}\,d\tw\ .
\eea
We distinguish $w$ and $\tw$ here because it will be necessary to
distinguish the Killing vectors $\partial_w$ and $\partial_{\tw}$
later on.
Now the metric \p{again} becomes
\be
ds^2_{11} =-\left({\rho^2\over Q} du\right)^2
+\left({Q}d\tw-{\rho^2\over Q}du\right)^2
 +{R^2\over 4}(\tsigma_3+{2\hat J\over R^2}\,d\tw)^2
 +{{R'}^2\over 4}({\tsigma}'_3+{2\hat J'\over {R'}^2}\,d\tw)^2 +\ldots
\label{againb}
\ee

In this form, the metric is still adapted for a Kaluza-Klein reduction
on $\tpsi$ (or $\tpsi'$). A reduction on $k=\partial_\tpsi=\partial_\psi$
proceeds exactly as in the previous section.\footnote{Actually,
at this point, we should note that the constraint
equation \p{newcons} is coordinate invariant, as well as
Lorentz invariant --- see below.} The only difference is that
an extra gauge field ${2\hat J/ R^2}\,d\tw$ appears on the $adS_3$ space.
However, this is trivial since it is a constant gauge field.

The reduction on
\be
k=\partial_{\tw}=\partial_w+{2\hat J\over R^2}\partial_\psi
+{2\hat J'\over {R'}^2}\partial_{\psi'}\ .
\label{newk}
\ee
turns out to be more interesting\footnote{The
Killing vector $k$ will have closed orbits
of radius $Q$ if the following identifications are made:
$(w,\psi,\psi')=(w+2\pi Q n_1, \psi+ 4\pi n_2 + {4\pi {\hat J} Q\over
R^2}n_1, \psi'+ 4\pi n_3 + {4\pi {\hat J}' Q\over
{R'}^2}n_1)$, where $n_i$ are integers.  
As a result note that the global geometry of the
unreduced space is no longer $adS_3\times S^3\times S^3\times\bE^2 $.}.
{}For a Kaluza-Klein reduction
on $\tw$, we must reorganize the metric \p{againb} into the
standard form. We do so by introducing new 1-forms
\bea
\hat e^u &= &{\rho^2\over\bQ} du-{\hat J\over 2\bQ }\ts_3
-{\hat J'\over 2\bQ }\ts'_3\nn
\hat e^w &=& \bQ d\tw - {\rho^2\over \bQ}du
+{\hat J\over2 \bQ }\ts_3+{\hat J'\over2 \bQ }\ts'_3\nn
\hat e^3 &=&{1\over2s}\tsigma_3\nn
\hat e^{3'}&=&{1\over2c}{\tsigma}'_3
\label{newframeb}
\eea
with which the metric \p{againb} may be written as
\be
ds_{11}^2=(\hat e^w)^2-(\hat e^u)^2+(\hat e^3)^2+(\hat e^{3'})^2+\ldots
\label{againc}
\ee
Here, $\bQ^2=Q^2+({\hat J/R})^2+({\hat J'/R'})^2$. The reduced
metric now comes from dropping $(\hat e^w)^2$ in \p{againc} above.
In this case, the remaining off-diagonal terms in $(\hat e^u)^2$
cannot be removed by a coordinate transformation, even locally.
Hence the reduced geometry is not a simple product of factors.
However, we will see that the structure of the isometry
supergoup is identical, discounting changes in purely bosonic
factors unrelated to Killing spinors, to that of the $J=J=0$
case, despite the fact that the metric has a direct product 
structure only in the $J=J' =0$ limit.

To determine the surviving Killing spinors, we need to determine
the constraint matrix $P$ in \p{newcons}. The simplest approach, here,
is to note that $P$ is coordinate
invariant and Lorentz covariant. Hence the constraint 
will be the same as that calculated for \p{newk} before any
change of frames and coordinates. {}Furthermore, since $P$ is linear in $k$
and since we saw
in the previous section that the constraints
for reducing on $\partial_w$, $\partial_\psi$ and $\partial_{\psi'}$
all coincided, precisely the same constraint arises here, namely,
$\Gamma_-\epsilon=0$.
Thus once again, the $\epsilon_+$ Killing spinors correspond to
supersymmetries in the reduced solution.

We must again be careful about the precise form of the Killing
spinors in the reduced theory, as the frame \p{frame} used in deriving
\p{resultb} can not be used after the reduction. One change which
must be accounted for is the introduction of \p{newframeb} for the
Kaluza-Klein reduction. A second slightly more subtle change comes
from the coordinate transformation \p{oldcoord} which introduces
various $\tw$ dependences which we have not explicitly accounted for.
{}First of all, in \p{resultb} and \p{OOmega},
one finds
\bea
\Omega(\psi,\theta,\phi)&=&\exp\left({\hat J\tw\over R^2}\Gamma^{21}\right)
\tOmega(\tpsi,\theta,\phi)\nn
\Omega'(\psi',\theta,\phi)&=&\exp\left({\hat J'\tw\over {R'}^2}\Gamma^{2'1'}
\right)\tOmega'({\tpsi}',\theta',\phi')\ .
\label{Omega}
\eea
Thus after the coordinate change the Killing spinors $\epsilon_+$
depend on $\tw$. However in the same way, $\tw$ now also appears
in the 1-forms: $e^1$, $e^2$, $e^{1'}$ and $e^{2'}$. {}For example,
\be
e^1={R\over2}\left(-\sin(\tpsi+{2\hat J\tw\over R^2})\, d\theta + 
\cos(\tpsi+{2\hat J\tw\over R^2})\sin\theta\, d\phi \right)\ .
\ee
Hence in the reduced theory, these 1-forms would be replaced by,
\eg, $\hat e^1$, $\hat e^2$, $\hat e^{1'}$ and $\hat e^{2'}$ defined using
the $\sigma$-forms defined using $\tpsi$ --- see \p{threesigma}.
As in the $\psi$ reduction above, these two sets of 1-forms are
related by rotations acting in the 1-2 and $1'$-$2'$ subspaces.
One finds
\be
\pmatrix{\hat e^1\cr\hat e^2}=L_{12}\pmatrix{e^1\cr e^2\cr}
\ee
with
\be
 L_{12}=
\pmatrix{\cos\left({2\hat J\tw\over R^2}\right)
             &\sin\left({2\hat J\tw\over R^2}\right)\cr
         -\sin\left({2\hat J\tw\over R^2}\right)
	     &\cos\left({2\hat J\tw\over R^2}\right)\cr}
=\exp\left[\pmatrix{0&1\cr -1&0\cr}{2\hat J\over R^2}\tw\right]
\label{rotateb}
\ee
and similarly for $L_{1'2'}$. The corresponding spinor
rotations \p{rotatespin} are then
\be
\tL_{12}=\exp\left(-{\hat J\tw\over R^2}\Gamma^{21}\right)
\qquad
\tL_{1'2'}=\exp\left(-{\hat J'\tw\over {R'}^2}\Gamma^{21}\right)
\label{rotatespinb}
\ee
Hence this rotation, which removes the $\tw$
dependence in the angular frames, at the same time removes
the $\tw$ dependence of the $\epsilon_+$ spinors (and
introduces it into the $\epsilon_-$). The net effect is
that in \p{resultb}, $\Omega\Omega'\rightarrow\tOmega\tOmega'$.

Now we also had to account for the change of frames \p{newframeb}.
Since both \p{frame} and \p{newframeb} describe the same metric,
they must be related by a Lorentz transformation.
{}First one finds that in this four-dimensional subspace
$\hat e^a=(L_4)^a{}_b\,e^b$ with
\be
L_4=\pmatrix{z&z-z^{-1}&-x/z&-y/z\cr
0&z^{-1}&x/z&y/z\cr
-x&-x&1&0\cr
-y&-y&0&1\cr}
\ee
where we have introduced the notation
\be
z={\bQ\over Q}\ ,\qquad x={\hat Js\over Q}\ ,\qquad
y={\hat J'c\over  Q }\ .
\label{notation}
\ee
The latter are not all independent, but rather from the definition of $\bQ$,
they satisfy the constraint: $z^2=1+x^2+y^2$.

To obtain the corresponding Lorentz transformation on the spinors
as in \p{rotatespin},
it is convenient to decompose $L_4$ as
$L_4=B\,W$ where $B$ is a boost
\be
B^a{}_b=\pmatrix{{1\over2}(z+z^{-1})&{1\over2}(z-z^{-1})&0&0\cr
{1\over2}(z-z^{-1})&{1\over2}(z+z^{-1})&0&0\cr
0&0&1&0\cr
0&0&0&1\cr}
\ee
and $W$ is the remaining transformation
\be
W^a{}_b=\pmatrix{1+{x^2+y^2\over2}&{x^2+y^2\over2}&-x&-y\cr
-{x^2+y^2\over2}&1-{x^2+y^2\over2}&x&y\cr
-x&-x&1&0\cr
-y&-y&0&1\cr}\ .
\ee
Now one finds $B=\exp(\omega_B)$ and $W=\exp(\omega_W)$ where
\bea
(\omega_B)_{ab}&=&\pmatrix{0&-\lambda&0&0\cr
                         \lambda&0&0&0\cr
			 0&0&0&0\cr
			 0&0&0&0\cr} \nn
(\omega_W)_{ab}&=&\pmatrix{0&0&x&y\cr
                          0&0&x&y\cr
			  -x&-x&0&0\cr
			  -y&-y&0&0\cr} 
\eea
with $\lambda=\log(z)$. 
Given these generators, we can write the
corresponding spinor transformations: $\tL=\tB\,\tW$ with
\bea
\tB&=&\exp\left({1\over2}\log z\Gamma^{wu}\right)={z^{1/2}}\Lambda_+
+{z^{-1/2}}\Lambda_-\nn
\tW&=&\exp \left((x\Gamma^{u3}+y\Gamma^{u3'})\Lambda_+\right)\nn
&=&1+(x\Gamma^{u3} + y\Gamma^{u3'})\Lambda_+ 
\eea
where we have defined the projection operators
\be
\Lambda_\pm={1\over2}(1\pm\Gamma^{wu})\ .
\label{projectwu}
\ee

So given the original Killing spinor solutions \p{resulta}
and \p{resultb},
they are transformed to the new frame  by
\be
\hat\epsilon=\tL_{12}\tL_{1'2'}\tL_4\,\epsilon=\tL_{12}\tL_{1'2'}
\tB\,\tW\,\epsilon
\label{transport}
\ee
Given that all of these transformation matrices commute with
$\Gamma_-$, the constraint $\Gamma_-\epsilon=0$ takes precisely the same
form on the $\hat\epsilon$ spinors. Hence the supersymmetries
of the solution in the reduced theory are given by $\hat\epsilon_+$.
Explicitly \p{transport} yields
\be
\hat\epsilon_+=
\left({1\over\rho} \left[z^{1/2}+z^{-1/2}(x\Gamma^{u3}+
 y\Gamma^{u3'})\right]
 -{2u\rho\over\sqrt{z} \g Q}\Gamma^{\rho u}\right)\tOmega\tOmega'
\chi_+^+ +{\rho\over\sqrt{z}}\,\tOmega\tOmega'\chi_+^-
\label{resultbb}
\ee
Note that the Killing spinor $\hat\epsilon^-_+$, which should
correspond to the supersymmetry which extends to the full rotating
solution in the reduced theory, still has essentially the same simple form
as with $J=J'=0$.

Having established the explicit form of the Killing spinors we deduce
that the full isometry superalgebra in the reduced spacetime
is now $D(2|1,\alpha)\times
U(1)\times U(1)$. Recalling eq. \p{fig} we see that
the effect of the rotation in each of
the two 4-planes is to break the extra bosonic $SU(2)$
rotational symmetries
of these 4-planes to $SO(2)\cong U(1)$, without affecting the
supersymmetry.

\section{Discussion}

We have found a new family of 1/8 supersymmetric rotating M-brane solutions,
with two independent rotation parameters. Many previous supersymmetric rotating
brane solutions can be found from the $\alpha\rightarrow 0$ limit of this new
solution by a combination of dualities and compactifications. The 
near-horizon limit of the new rotating brane solution is the 1/2
supersymmetric $adS_3\times S^3\times S^3\times \bE^2$ vacuum, irrespective
of the rotation parameters (within the limits for which there exists a
non-singular event horizon). There is thus a fourfold increase of
supersymmetry near the horizon, although this is invariably reduced to
a two-fold increase on $S^1$ compactification by the identifications required
to perform the reduction. Non-trivial $S^1$ compactifications lead to a 
replacement of $adS_3$ by $adS_2$ or $S^3$ by $S^2$ in the near-horizon limit.
The possible near-horizon geometries obtainable this way were classified in
\cite{boonstrab} for non-rotating intersecting branes; we now see that
the same results apply in the rotating case, at least locally. 

{}Finally we note that the $(1|MW,M2,M5,M5)$ configuration of M-theory has
a IIB dual as $(1|IIW,D1,D5,D5)$ so that the entropy associated with the event
horizon is expected to correspond to a counting of D-brane microstates along
the lines of \cite{microstates}. It would be of interest to see how the
rotation affects these calculations. We leave this to future investigation.

\vskip 1cm

\noindent{\bf Acknowledgements:} JPG would like to thank the ITP 
for hospitality during the workshop ``Dualities in String Theory''
where this work began, and the EPSRC for partial support. RCM was supported
in part by NSERC of Canada and {}Fonds FCAR du Qu\'ebec. 
Research at the ITP
was supported by NSF Grant PHY94-07194.
 

\end{document}